\documentclass{aa}
\usepackage{hyperref}
\usepackage{txfonts}
\usepackage{graphicx}
\usepackage{amssymb}
\usepackage{amsmath}
\usepackage{natbib}

\def\gr{$\gamma$-ray}

\begin{document}
\title{Search for spectral features in extragalactic background light with gamma-ray telescopes}
\author{A. Korochkin \inst{1,2} \and 
A. Neronov \inst{1,3} \and
D. Semikoz \inst{1} }
\date{}
\institute{APC, Universite Paris Diderot, CNRS/IN2P3, CEA/IRFU \email{alexander.korochkin@apc.in2p3.fr} \and Institute for Nuclear Research of the Russian Academy of Sciences, 60th October Anniversary st. 7a, 117312, Moscow, Russia \and Astronomy Department, University of Geneva, Ch. d'Ecogia 16, 1290, Versoix, Switzerland}
\label{firstpage}

\abstract
{
Cumulative optical and infrared emission from galaxies accumulated over cosmological time scales, the extragalactic background light (EBL), could be probed by complementary techniques of direct observations and source counting in the visible and infrared as well as via its imprint on the signal of distant active galactic nuclei in gamma rays. 
}
{
We compare the visible and infrared measurements with the gamma-ray constraints and study if the discrepancies of the measurements with different methods could be due to the presence of features in the EBL spectrum that are localised in the micron wavelength range.
} 
{
We combined data on time-averaged spectra of selected blazars that were obtained by Fermi and ground-based \gr\ telescopes. We also modelled the effect of absorption on EBL while allowing for the existence of a previously unaccounted spectral feature.
}
{ 
We show that a previously reported 'excess' in EBL flux in the $\sim 1$~micron wavelength range is consistent with gamma-ray measurements, that is, if the excess has the form of a narrow feature of the width $\delta \lambda < \lambda$ and an overall flux of up to 15 nW/(m$^2$ sr) above the 'minimal' EBL, which is estimated from the visible and infrared source counts. Such 'bump-like' spectral features could originate, for example, from decaying dark-matter particles, or either axions or peculiar astrophysical processes in the course of star-formation history. We discuss the possibilities for the search of spectral features in the EBL with the Cherenkov Telescope Array (CTA). 
}
{}


\maketitle
\section{Introduction}

The spectrum of the extragalactic background light (EBL) in the visible and near-infrared band encodes valuable cosmological  information on the history of the formation of stars and galaxies and possibly on other astrophysical processes, which have resulted in visible light emission over the course of the history of the Universe. Precision measurements of its properties are, however, challenging because of the presence of zodiacal light background (see \citep{dwek} for a review). To handle this problem, a variety of indirect techniques were developed. On the one hand, robust lower constraints come from the deep field observations of the Universe. This approach uses the method of galactic counts and lower limits on the EBL intensity provided by a sum of contributions from resolved sources \citep{Xu:2004zg,Madau:1999yh,Keenan:2010na,Fazio:2004kx}.   

On the other hand, a complementary approach for the measurement of the EBL is based on the absorption of gamma-rays via the pair production off EBL photons \citep{Ahnen:2016gog,Abramowski:2012ry,desai,Ajello:2018sxm,Acciari:2019zgl}. This effect leads to the distance-dependent suppression of the \gr\ flux from extragalactic sources at the highest energy. However, the precision of the measurements of the EBL using the high-energy suppression suffers from the uncertainty of the knowledge of the intrinsic primary source spectrum. Furthermore, \gr\ measurements do not provide a measurement of the EBL flux at a particular wavelength. The wavelength resolution of the measurements is limited by the width of the pair production cross-section that peaks at the centre of mass energy, which is somewhat above the threshold $E_{thr}=2mc^2$ of  twice the rest energy of the electron and decreases as $E^{-1}\mbox{ln}(E)$ at $E\gg E_{thr}$. Measurements of the EBL from the \gr\ data conventionally adopt assumptions about the shape of the EBL spectrum. 

\begin{figure}
                \resizebox{\hsize}{!}{\includegraphics{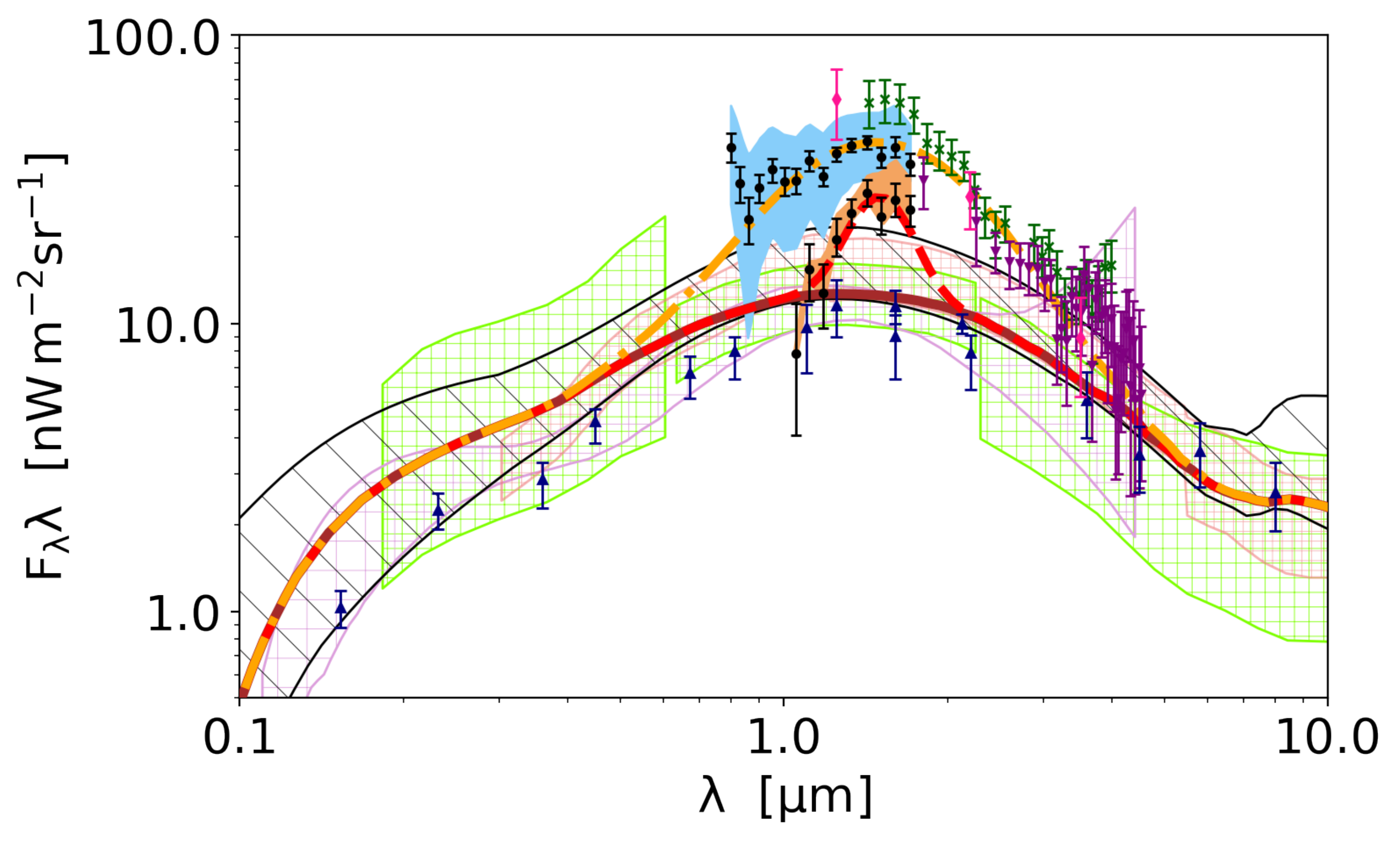}}
                \caption{ SED of the EBL obtained by various methods. 
                                 {\it Direct measurements:} The purple arrows are results from AKARI
                                 \citep{Tsumura:2013iza}, while green asterisks are from the 
                                 reanalysis of IRTS \citep{Matsumoto:2015fma}. Pink diamonds are
                                 from the reanalysis of COBE/DIRBE measurements \citep{Sano:2015bsa}
                                 \citep{Sano:2015jih}. Black data points together with blue and orange
                                 systematic uncertainty are derived from CIBER \citep{Matsuura:2017lub}
                                 and correspond to nominal and minimum EBL models.
                                 {\it Lower limits:} Dark blue upward arrows combine EBL lower limits
                                 obtained by different experiments: GALEX \citep{Xu:2004zg}, Hubble 
                                 Deep Field \citep{Madau:1999yh}, Subaru \citep{Keenan:2010na}, and
                                 Spitzer/IRAC \citep{Fazio:2004kx}. {\it EBL from $\gamma$-ray 
                                 absorption:} Striped lime, red, and purple bands are from MAGIC 
                                 \citep{Acciari:2019zgl}, HESS \citep{Abramowski:2012ry}, and Fermi/LAT
                                 \citep{Ajello:2018sxm}, correspondingly.
                                 {\it Modelling:} Dark red solid line is for baseline EBL model of 
                                 \citet{Gilmore:2011ks}. Black striped band shows the allowed range of EBL models 
                                 obtained with the global fit to  EBL  by
                                 \citet{Korochkin:2017plb}. Dash-dotted orange and dashed red lines
                                 are additional wide and narrow Gaussian components.}
                \label{fig:ebl}
        \end{figure}

The first direct detection of the EBL was carried out by the Cosmic Background Explorer (COBE) \citep{Wright:1999zb} and Infrared Telescope in Space (IRTS) \citep{Matsumoto:2004dx} (restated in \citep{Matsumoto:2015fma}). After subtraction of zodiacal light and the contribution of the galaxy, both experiments reported results with a peak intensity of 60 $nW/m^2/sr$ at about 1.5 $\mu m,$ which is approximately five  times higher than the corresponding intensity from galaxy counts. The shape of the detected 'near-IR excess' has a sharp cutoff at 1 $\mu m$ and a long tail towards longer wavelengths. Subsequent measurements by the Cosmic Infrared Background Experiment (CIBER) \citep{Matsuura:2017lub} confirm the presence of an 'excess' but with a lower intensity of 42.7 $nW/m^2/sr$ using Kelsall ZL model \citep{Kelsall:1998bq} (CIBER 'nominal' EBL) and 28.7 $nW/m^2/sr$ for the model independent analysis (CIBER 'minimal' EBL). Recent results from AKARI \citep{Tsumura:2013iza} between $\sim 2 \mu m$ and $\sim 5 \mu m$ also demonstrate an increase in EBL intensity at shorter wavelengths. The situation remains uncertain: The direct measurements are systematically above lower bounds (see Fig. \ref{fig:ebl}). This discrepancy is interesting because it might point to the presence of a truly diffuse emission component that is not resolvable into point sources, which, for example, could arise from interactions (annihilation or decay) of dark-matter particles in the Milky Way dark-matter halo and the cumulative signal from dark-matter interactions in the halos of all galaxies accumulated over the cosmological time scale.

It is clear from Fig. \ref{fig:ebl} that direct measurements of the EBL are in tension with the data from the \gr\ measurements in the wavelength range between 1~$\mu$m and 4~$\mu$m. It is possible that the precision of one of the direct measurements suffers from some unaccounted systematic effect that makes the \gr\ measurements more reliable \citep{dwek05,kawara17}. Nevertheless, measurements of the EBL from the \gr\ data conventionally adopt assumptions about the smooth shape of the EBL spectrum and they are thus insensitive to narrow spectral features. Despite the fact that it was shown that IRTS excess is unlikely to be of extragalactic origin since it leads to an unphysical upward break in the intrinsic spectrum of the blazar PKS 2155-304 \citep{Dwek:2005sp}, there is still a possibility that the \gr\ and direct measurements are in fact consistent.\ This would be the case once the model of the EBL used in the \gr\ data analysis is adjusted to take into account a more sophisticated shape of the EBL. This could be the case if, for example, the EBL has an excess flux (a 'feature') concentrated in a narrow wavelength range, which has not been accounted for in the \gr\ analysis. Examples of such a relatively narrow feature is shown by the red dashed line in Fig. \ref{fig:ebl}. 

Such features could originate from processes that have a characteristic wavelength or energy scale. Examples are given by the emission from Population III stars \citep{populationIII}, which produce strong emission at the wavelength of Ly alpha lines redshifted towards near infrared, hypothetical decaying particles \citep{kohri17}, and axion-like particles \citep{axions} that are expected to produce a feature at the energy close to the particle rest energy.

In what follows, we explore constraints on the narrow spectral features in the EBL spectrum imposed by existing \gr\ data. We show that in spite of limited energy and wavelength resolution of the \gr\ measurement technique, the data impose constraints on the position, width, and overall flux of the features. For a given reference wavelength of the feature, the \gr\ data provide constraints on the flux as a function of the spectral width of the feature. We show that the minimal EBL measurement by CIBER, which is equivalent to a narrow feature superimposed onto the overall low EBL flux at the level of the lower bounds from galaxy counts, is consistent with the \gr\ data; whereas nominal EBL by CIBER is ruled out. 

Starting from \citep{Franceschini2008}, EBL models are conventionally in agreement with gamma-ray constraints. \citep{Korochkin:2017plb} calculated the allowed range of EBL models by combining all observations except gamma-ray constraints, which include the measurements discussed above (e.g. star formation rate, etc). The models of \citep{Finke:2009xi, Dominguez:2010bv,Gilmore:2011ks,Franceschini:2017iwq, Andrews:2017ima} are located within this allowed range. In the range around 1 micron, those models are close to the lower bound of the allowed band, which is in agreement with recent gamma-ray constraints by Imaging Atmospheric Cherenkov Telescopes (IACTs) HESS \citep{Abramowski:2012ry} and MAGIC \citep{Acciari:2019zgl}, as well as the space telescope Fermi/LAT \citep{Ajello:2018sxm}. The analysis reported in the following sections uses the model of \citep{Gilmore:2011ks} as the baseline that is roughly at the level of the low bound on the EBL.

\section{Combining Fermi/LAT and IACT spectra}

Our analysis relies on the combined time-averaged Fermi/LAT and IACT spectra of blazars. Such combined spectra could only be produced for sources that were the subject of long, multi-year monitoring observations with IACTs. This is not the case for most of the TeV detected blazars because ground-based telescopes typically observe these sources during specific activity periods. 
        
To define a set of sources for which the production of time-averaged combined Fermi/LAT and IACT spectra is possible, we performed source selection in the following way. The initial selection was done from the TeVCat online source catalogue \citep{TeVCat} \footnote{Available at \url{http://tevcat.uchicago.edu}}, which aggregates observations performed by different observatories. The TeVCat catalogue includes 71 sources that are classified as blazars, with the following sub-classes: 'HBL', 'IBL', 'LBL', 'FSRQ', 'BL Lac (class uncertain)', and 'Blazar'. Out of the 71 sources from the TeVCat, we only retained sources with spectroscopically measured redshift, derived from emission lines. This restricts the list of blazars to 42 sources. 

Then, we imposed a requirement that the selected blazars should have been observed for at least two years during the period of operation of the Fermi telescope, that is, after 2008 and the reported spectrum should be averaged over the time of observation. IACTs only typically observe sources for several dozens of hours per year during several months of observations. The preference of only observing the blazar during flaring activity introduces a bias in the flux and spectral measurements. To avoid possible effects of this kind of bias, we selected only those sources for which long-term exposures over several years have been reported in the literature. In case the long-term observation data were published in the form of low-flux-state and high-flux-state spectra, we used low-state ones. If the spectra were not divided on low- and high-states, we checked that the blazar did not exhibit strong flares during the IACT observational period, when the flux increased more than ten times as compared to the average. After imposing this constraint, 21 blazars have been left in the sample.

We have further imposed a bound on the source flux $F>0.03 F_{Crab}$, where $F_{Crab}$ is the flux of the Crab Nebula in order to assure sufficient quality of the gamma-ray spectral measurements. This leaves five blazars in our source sample. In addition, we added PKS 2005-489 and PKS 2155-304, which satisfy all our criteria, but they were observed before 2008. All the blazars selected for the analysis are listed in Table \protect\ref{table:bl_list}.

\begin{table*}
        \centering
        \caption{Sample of seven blazars, selected for analysis.}
        \label{table:bl_list}
        \begin{tabular}{l l l l l l l l}
                \hline\hline
                Name             & Ra     & Dec   & z     & Flux (Crab) & Instrument & Obs. period        & Reference      \\ [0.5ex]
                \hline
                1ES 1011+496 & 153.76 & 49.43 & 0.212 & 0.05 & MAGIC & 2011-2012 & (1) \\
                1ES 1215+303 & 184.45 & 30.10 & 0.131 & 0.035 & MAGIC & 2010-2011 & (2) \\
                1ES 1218+304 & 185.36 & 30.19 & 0.182 & 0.08 & VERITAS & 2008-2013 & (3) \\
                1ES 1959+650 & 299.99 & 65.14 & 0.048 & 0.64 & VERITAS & 2007-2011 & (4) \\
                PKS 1510-089 & 228.21 & -9.10 & 0.361 & 0.03 & MAGIC & 2012-2017 & (5) \\
                \hline
                PKS 2005-489 & 302.36 &-48.83 & 0.071 & 0.03 & HESS     & 2004-2007 & (6) \\      
                PKS 2155-304 & 329.72 &-30.22 & 0.116 & 0.15 & HESS     & 2005-2007 & (7) \\
                \hline
        \end{tabular}
        \tablefoot{Ra and Dec are equatorial coordinates of the blazar, z is the redshift, and flux represents the power of the source, measured with the IACT.} 
        \tablebib{(1)~\citep{Aleksic:2016wfj}; (2) \citep{Aleksic:2012}; (3) \citep{Madhavan:2013sea}; (4) \citep{Aliu:2013nza}; (5) \citep{Acciari:2018}; (6) \citep{Acero:2009}; (7) \citep{Abramowski:2010}.}
\end{table*}

\subsection{Fermi/LAT data analysis}

For each selected blazar from Table \ref{table:bl_list}, we calculated the time-averaged spectrum using Fermi/LAT data. The Fermi/LAT  data that were collected during the ten-year-long Fermi/LAT mission from August 4, 2008 to September 19, 2018 were processed using Fermi Science Tools software version 1.0.2. We used Fermi Pass 8 Release 3 data with 'SOURCE' class events from both the front and back. The Galactic interstellar emission model used in the analysis was 'gll\_iem\_v07.fits'. The isotropic background model was  'iso\_P8R3\_SOURCE\_V2\_v1.txt'. The instrument response functions were  'P8R3\_SOURCE\_V2'. Final spectra were obtained using Fermi likelihood analysis with the help of the {\it gtlike} routine. We imposed standard cuts for the Fermi/LAT unbinned likelihood analysis, such as Earth zenith angle $< 90^{\circ}$. Source models were created on the basis of the fourth Fermi/LAT Source Catalogue 4FGL \citep{Fermi-LAT:2019yla}. The spectra were calculated in ten energy bins covering energy range from 1 GeV up to 1 TeV. To account for the systematic uncertainty of the effective area, we added 10\% errors to every bin.

        \begin{figure*}
                \centerline{%
                        \includegraphics[width=0.5\linewidth]{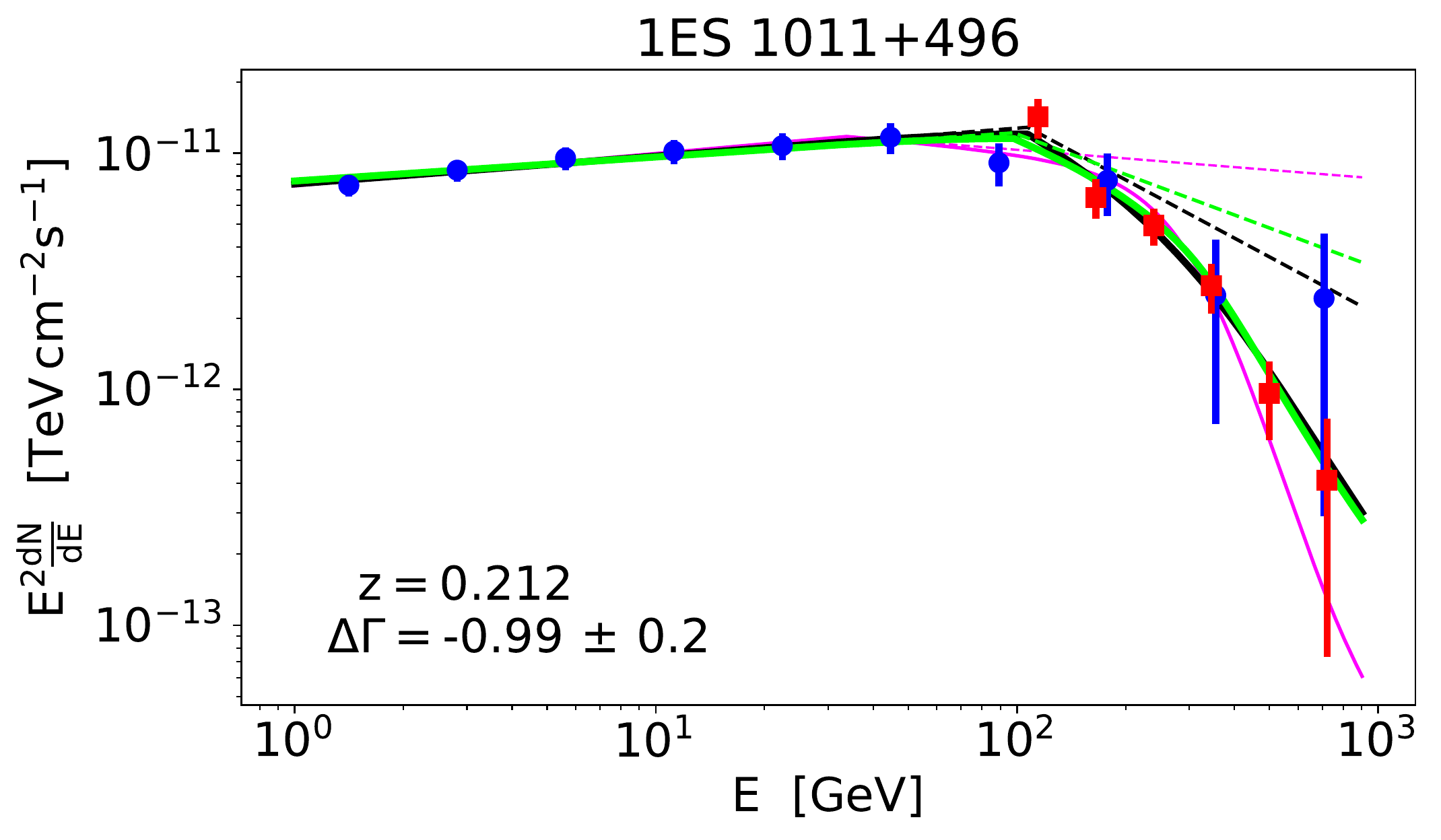}
                        ~~~~
                        \includegraphics[width=0.5\linewidth]{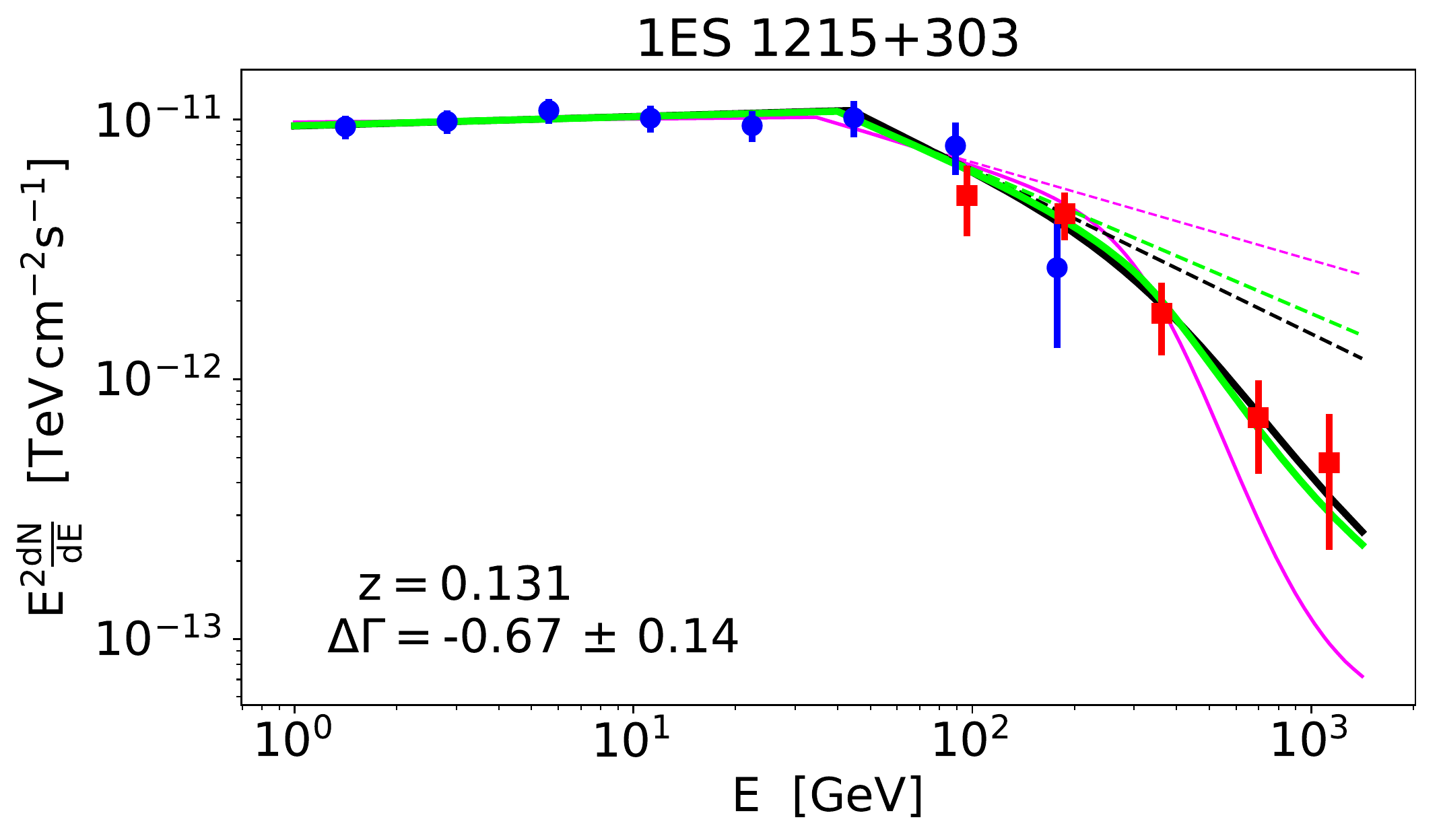}
                }
                \bigskip
                \centerline{%
                        \includegraphics[width=0.5\linewidth]{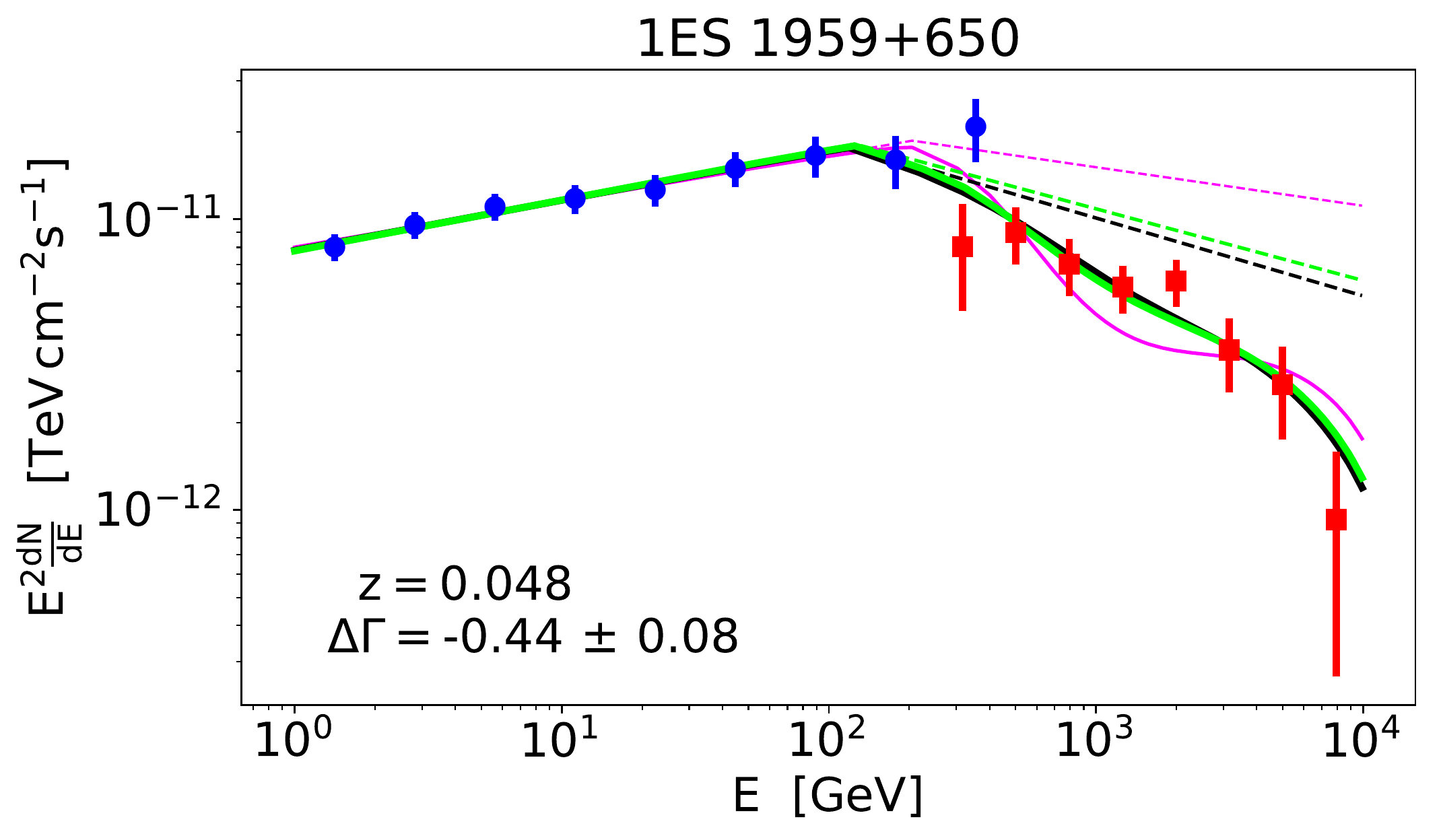}
                        ~~~~
                        \includegraphics[width=0.5\linewidth]{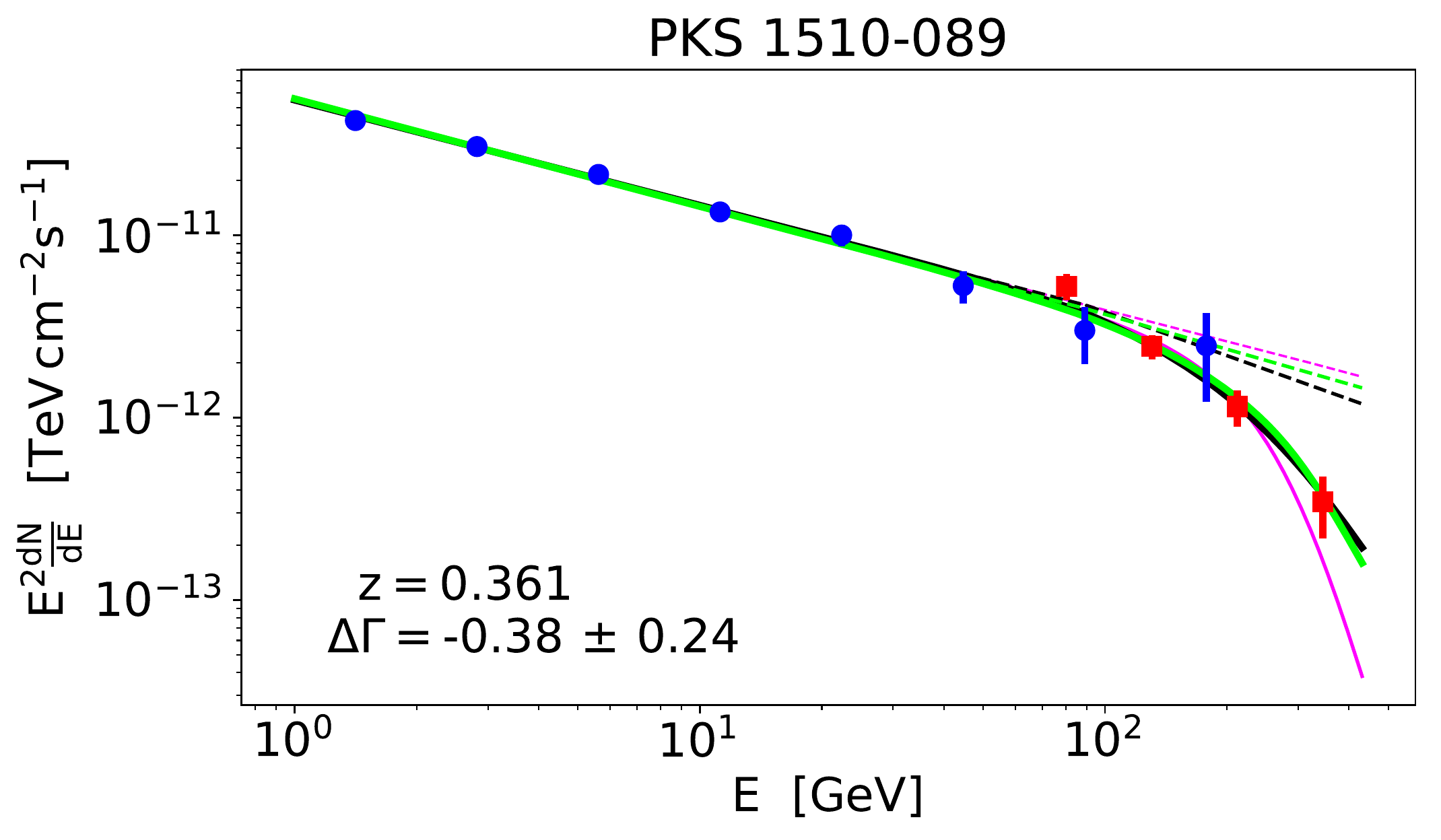}
                }
                \bigskip
                \centerline{%
                        \includegraphics[width=0.5\linewidth]{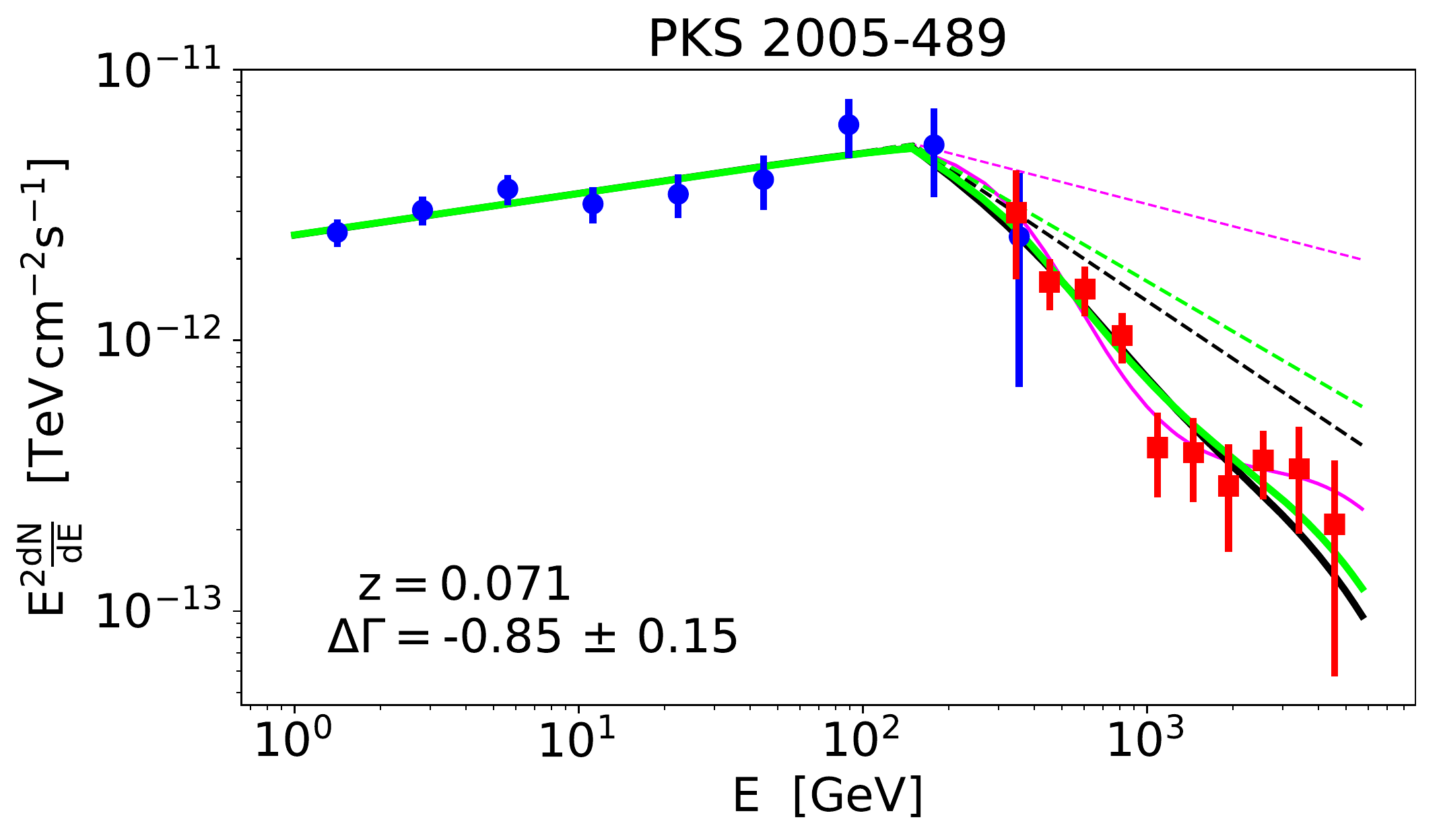}
                        ~~~~
                        \includegraphics[width=0.5\linewidth]{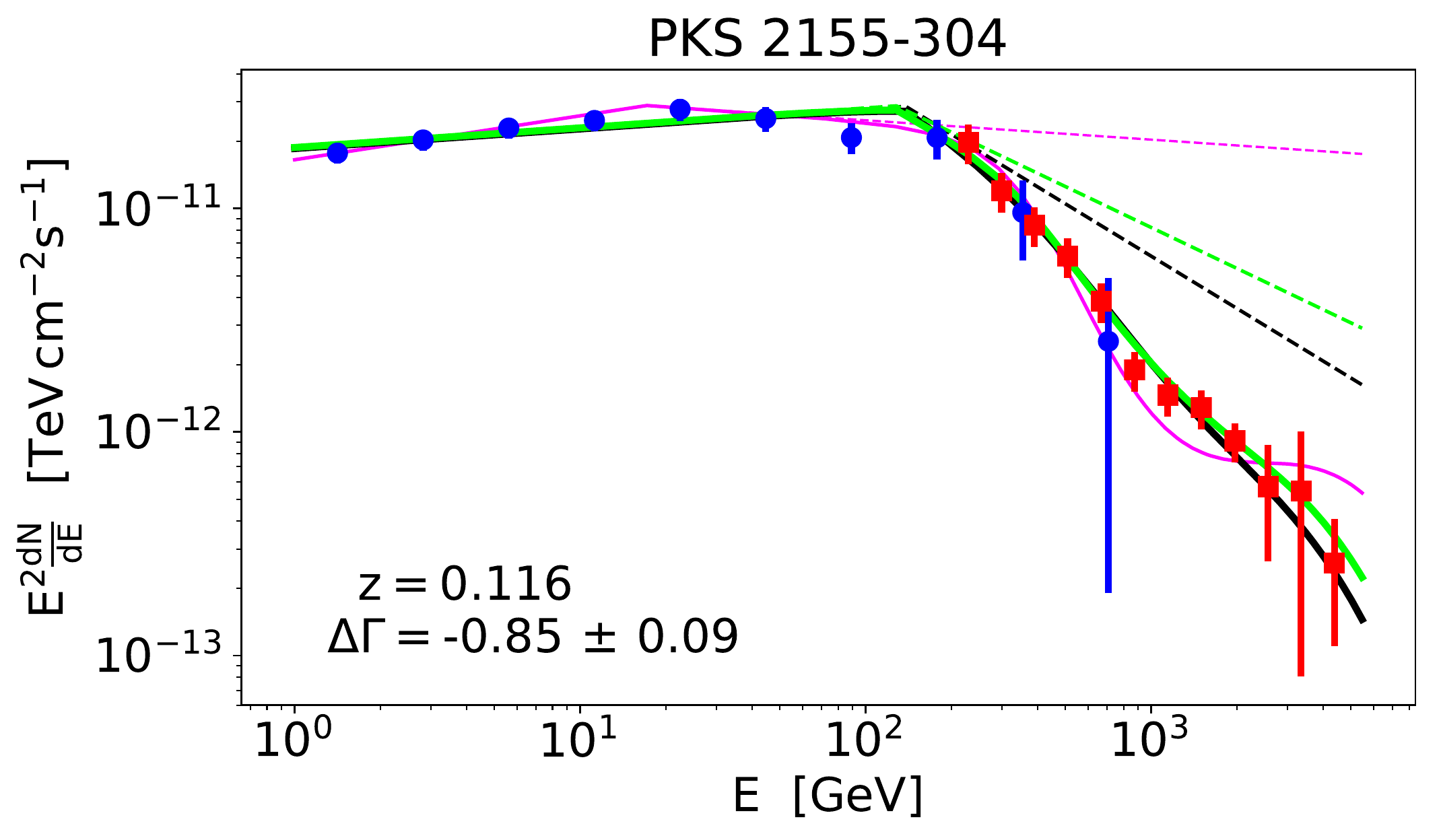}
                }
                \centerline{%
                        \includegraphics[width=0.5\linewidth]{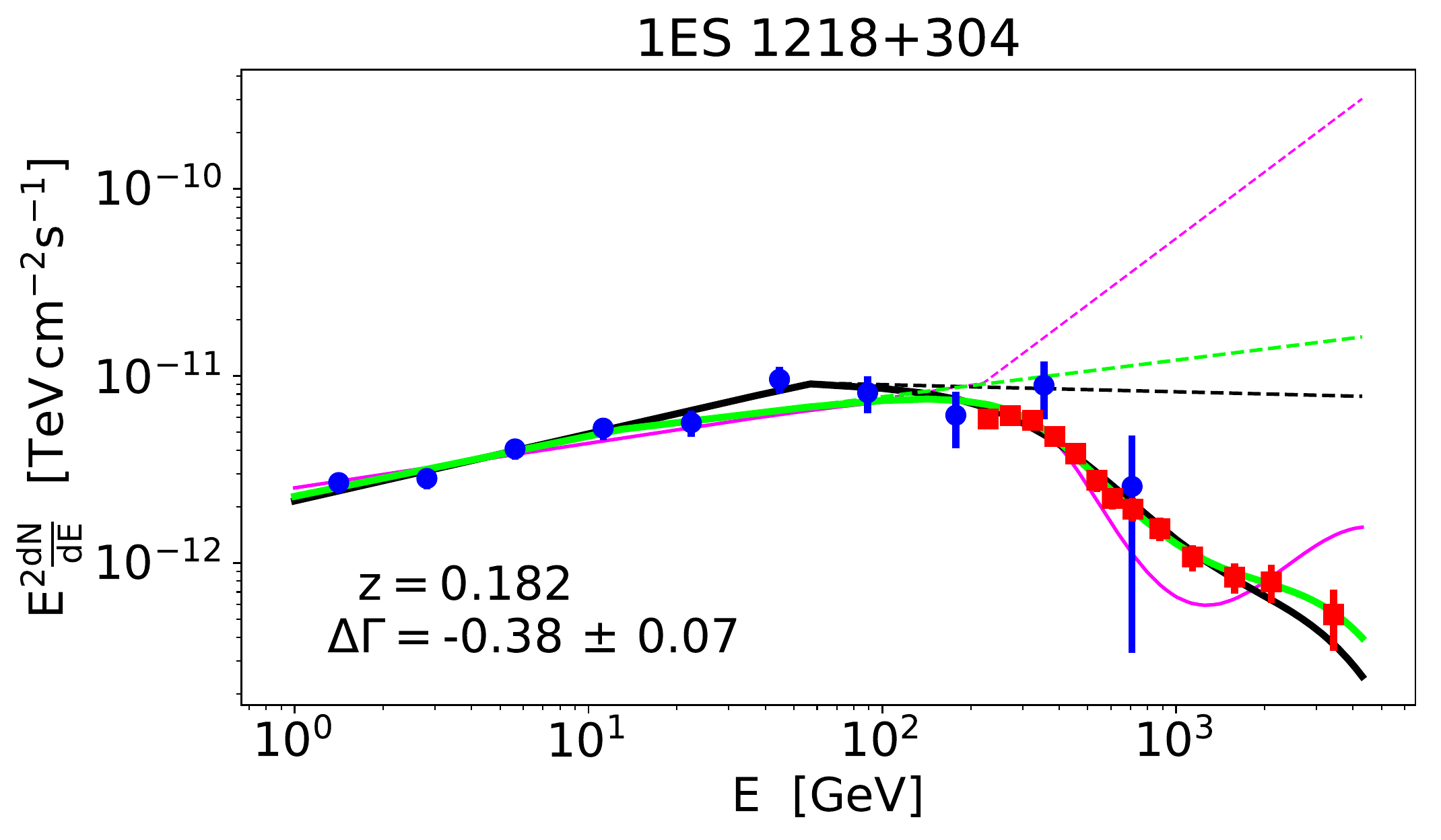}
                }
                \caption{ Broad band blazar's spectra studied in this work.
                                 Blue circles and red squares correspond to Fermi/LAT and IACT parts
                                 of the spectra. Black dashed thin and black solid thick lines are the best-fit broken 
                                 power law intrinsic spectra and observed spectra absorbed with the 
                                 baseline EBL model. Dashed and solid green lines indicate the same thing, but
                                 for the baseline EBL with an additional narrow bump at the level of the 
                                 minimal EBL excess measured by CIBER \citep{Matsuura:2017lub}.
                                 Dashed and solid magenta lines correspond to the baseline EBL with 
                                 additional nominal EBL excess detected by CIBER \citep{Matsuura:2017lub}.}
                \label{fig:specs}
        \end{figure*}

        \begin{figure}
                \resizebox{\hsize}{!}{\includegraphics{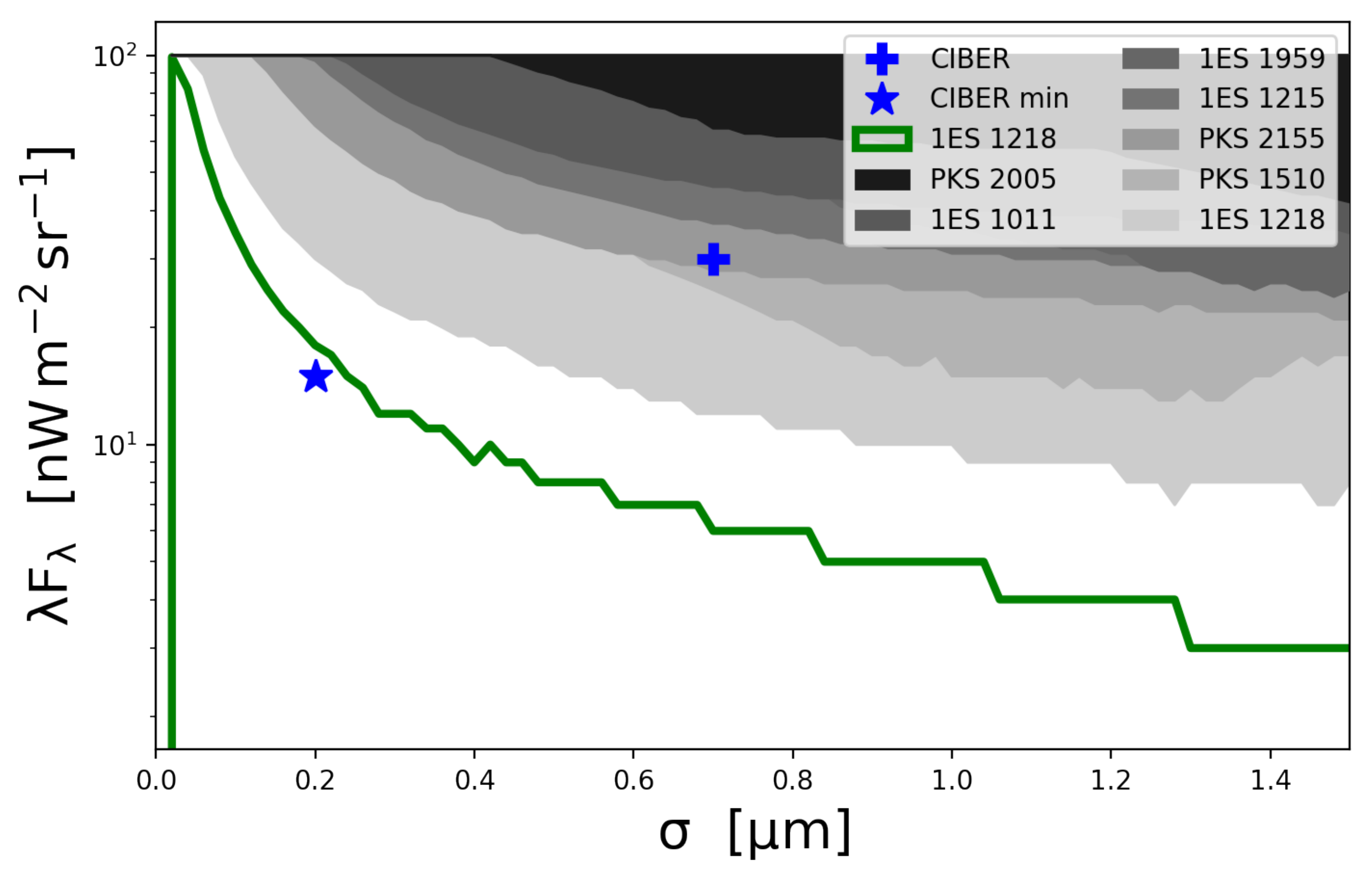}}
                \caption{Cross section of three dimensional parameter space calculated for the bump position $\mu=1.5\ \mu$m. Grey areas correspond to the cross sections of 95\% confidence level constraints on the parameters of the log-Gaussian EBL bump, obtained with different blazars. The strongest constraint is imposed by 1ES 1218+304 data. For this source, the cross section of 68\% confidence contour around the best-fit value is also shown. Blue cross in the excluded region and the star within the 68\% confidence contour show the two bump measurements reported by CIBER \citep{Matsuura:2017lub}.}
                \label{fig:ebl_bplaw} 
        \end{figure}

        \begin{table*}
                \centering
                \caption{Fitting parameters for each blazar.}
                \label{table:bl_params}
                \begin{tabular}{l l l l l l l l}
                \hline\hline
Name             & $\Gamma_1$           & $\Delta\Gamma$        & $E_{br}$ & $\chi^2$      & $\chi^2/d.o.f.$ & $\Delta\chi^2$ & $\chi^2_{b}/d.o.f$          \\
 & & & GeV & & & & \\
\hline
1ES 1011+496 & 1.89$\pm$0.04    & -0.99$\pm$0.20        & 107$\pm$24    &       6.103         & 0.469 &       -0.291  & 0.447 \\
1ES 1215+303 & 1.98$\pm$0.05    & -0.67$\pm$0.14        & 44$\pm$13             &       3.725   & 0.373 & +0.449  & 0.417 \\
1ES 1218+304 & 1.65$\pm$0.05    & -0.38$\pm$0.07        & 57$\pm$19             &       12.622         & 0.631 &       -1.395  & 0.561 \\
1ES 1959+650 & 1.83$\pm$0.04    & -0.44$\pm$0.08        & 119$\pm$48    &       9.120   & 0.651 &   +0.918  & 0.717 \\
PKS 1510-089 & 2.56$\pm$0.04    & -0.38$\pm$0.24        & 90$\pm$50             &       4.322   & 0.480 &   -0.010  & 0.479 \\
\hline
PKS 2005-489 & 1.86$\pm$0.05    & -0.85$\pm$0.15        & 150$\pm$48    &       10.671  & 0.667   &       -0.942  & 0.608 \\ 
PKS 2155-304 & 1.93$\pm$0.03    & -0.85$\pm$0.09        & 138$\pm$32    &       10.843  & 0.571   &       -1.542  & 0.489 \\
                \hline
                \end{tabular}
                \tablefoot{$\chi^2/d.o.f.$ represents the quality of the fit with the baseline EBL model while $\chi^2_{b}/d.o.f$ is the quality of the fit for the EBL with an additional narrow bump, corresponding to the minimal EBL excess measured by CIBER \citep{Matsuura:2017lub}, $\Delta\chi^2 = \chi^2_{b} - \chi^2$.}
        \end{table*}

\subsection{Spectral fitting}

The effect of absorption by EBL is imprinted on intrinsic source spectrum. We find that a broken powerlaw model provides a satisfactory description of the observed spectra of the selected blazars. The model has four parameters: normalisation $A$, powerlaw indices $\Gamma_1$ and $\Gamma_2 = \Gamma_1 - \Delta\Gamma,$ and the break energy $E_{br}$:        \begin{equation}
                \begin{cases}
                        F_0(E) = A \left(E/E_{br}\right)^{-\Gamma_1}~~~~~\,\mbox{for}~E < E_{br}\\
                        F_0(E) = A(E/E_{br})^{-\Gamma_2}            ~~~~~~\,\, \mbox{for}~E > E_{br}\\
                \end{cases}
        .\end{equation}
The observed spectrum is
        \begin{equation} \label{bl_spec}
                F(E) = F_0(E)\,e^{-\tau(E, z)}
        ,\end{equation}
where $\tau$ is the energy-dependent optical depth for the pair production on EBL for a source at the redshift $z$. 

The next step is to choose the EBL model to absorb intrinsic spectra. We chose the model of \citep{Gilmore:2011ks} as it is one that has the lowest optical depth. This model has a nearly identical EBL spectrum to that of the model of \citep{Franceschini2008} in the visible and near-infrared range at $z=0$.  The results of our study are not significantly impacted by choosing this particular EBL model, which has the near-infrared and visible band flux at the level that is close to the lower bound derived from the galaxy counts at $z=0$. 
        
The spectra of the seven selected sources fitted with the spectral model described above are shown in Figure \protect\ref{fig:specs}. One can see that Fermi/LAT and IACT parts of the spectra are in good agreement with the fitted broken powerlaw model with $\chi^2/d.o.f. < 1$ for each blazar.

\subsection{Modelling of infrared and visible excess}

A good fit to the data is provided by the broken powerlaw model modified by the effect of absorption on 'low-flux' EBL at the minimal possible level estimated from galaxy counts. Although one could notice that all the spectral fits include a break in the intrinsic source spectra at around 100 GeV, that is, at the energy level above which the EBL effect becomes important. This hints to the possibility that, in fact, a higher level of EBL could be accommodated by the spectral fits. Weaker spectral break in the intrinsic spectrum could be compensated by stronger absorption on the EBL. 

To explore this possibility, we introduced an excess in the EBL spectrum, as a broad spectral feature localised around a characteristic energy. Such a spectral feature could be generated, for example, by a specific star-formation history (Population III stars \citep{populationIII}) or by the photons injected by interactions of exotic decaying particles \citep{kohri17}, that is, axion-like particles \citep{axions}) of the mass close to 1 eV. 

We modelled the additional excess EBL component with the log-Gaussian spectral shape $B_{\lambda}(\lambda)$:

\begin{equation} 
        \label{gauss_bump}
        \lambda B_{\lambda} = B\, \exp\Big(-\frac{\log^2(\lambda / \mu)}
        {2\,\log(1 + \frac{\sigma}{\mu})^2}\Big)
,\end{equation}  
as shown in Fig. \ref{fig:ebl}. This function appears as a Gaussian when plotted on a logarithmic scale (by analogy with log-parabola). It has three parameters: normalisation $B$, central wavelength $\mu,$ and  width $\sigma$. Figure \protect\ref{fig:ebl} shows that such a log-Gaussian can fit both the 'broad' excess at the level of the highest direct EBL measurements and the EBL excess reported by CIBER \citep{Matsuura:2017lub}. For the former, we should take $B = 30 \,nW/m^2/sr$ and $\sigma = 0.7 \,\mu$m, while for the latter $B = 15 \,nW/m^2/sr$ and $\sigma = 0.2 \,\mu$m. Both bumps are positioned at $\mu = 1.5 \,\mu$m.

To explore what kind of excess EBL flux could be accommodated to in the $\gamma$-ray data, we scanned over the height, width, and position of the bump. We also studied how the quality of the fits to the gamma-ray spectra improves (worsens) with changes in the parameters. At the first step, we defined a vector $\theta$ as a set of parameters $\theta = (B, \mu, \sigma),$ and we calculated $\chi^2_i(\theta)$ and the parameters $\Gamma_{1,i}(\theta)$, $\Gamma_{2,i}(\theta),$ and $E_{br,i}(\theta)$ of the fit of the $i$-th blazar for every point $\theta$ in range $0<B<100  \,nW/m^2/sr$ , $0.1<\mu<10  \,\mu m$, $0 < \sigma < \, 1.5 \,\mu$m, sampling the parameter space at small steps $\delta\theta$. Then, we imposed a cut on the hardness of the intrinsic blazar spectra and removed all the points $\theta$ where $\Gamma_{2, i} < 1.5$ for at least one blazar.

Thereafter, we derived the 2-$\sigma$ excluded region of parameter space for each blazar. The point $\theta$ is excluded if $\chi^2_i(\theta) > \chi^2_{i,0} + 8.02$ where $\chi^2_{i,0} $ corresponds to the best fit without additional excess and the value $8.02$ corresponds to the 2-$\sigma$ inconsistency level for the nested models with three added parameters. The 1-$\sigma$ preferred regions are those where the condition $\chi^2_i(\theta) < \chi^2_{i, min}+ 3.53$ is satisfied and $\chi^2_{i, min}$ corresponds to the best fit of the model with the additional excess.

The results are presented in Figures \protect\ref{fig:ebl_bplaw} and \protect\ref{fig:ebl_bplaw1}. The first plot is the cross-section of the three dimensional parameter space at $\mu=1.5\ \mu$m. It shows an upper bound on the normalisation of the excess as a function of its spectral width. The strongest constraints come from the blazar 1ES~1218+304. In the limit $\sigma > 1 \mu$m, the width of the bump becomes large enough to correspond to the change of the overall normalisation of the EBL. In this case we have found an upper bound
\begin{equation}
        B < 10\mbox{ nW/(m}^2\mbox{sr)}
\end{equation}
at 2-sigma level. This result is consistent with similar constraints derived by HESS \citep{Abramowski:2012ry} and MAGIC \citep{Ahnen:2016gog}.

The upper bound on the excess  EBL flux normalisation is relaxed if the excess feature is narrow, with the width $\sigma \ll 1\ \mu$m. In particular, Fig. \ref{fig:ebl_bplaw} shows that the narrow excess with the flux level comparable to the CIBER minimal EBL is not ruled out by the $\gamma$-ray observations while the CIBER nominal EBL is strongly excluded. In fact, by fitting the data of 1ES~1218+304 with a model EBL with a narrow bump, we find that the best fit is not achieved with a model without the bump, but with a bump which is stronger and narrower than CIBER minimal.

\subsection{Search for spectral features in the EBL spectrum}

\begin{figure}
                \resizebox{\hsize}{!}{\includegraphics{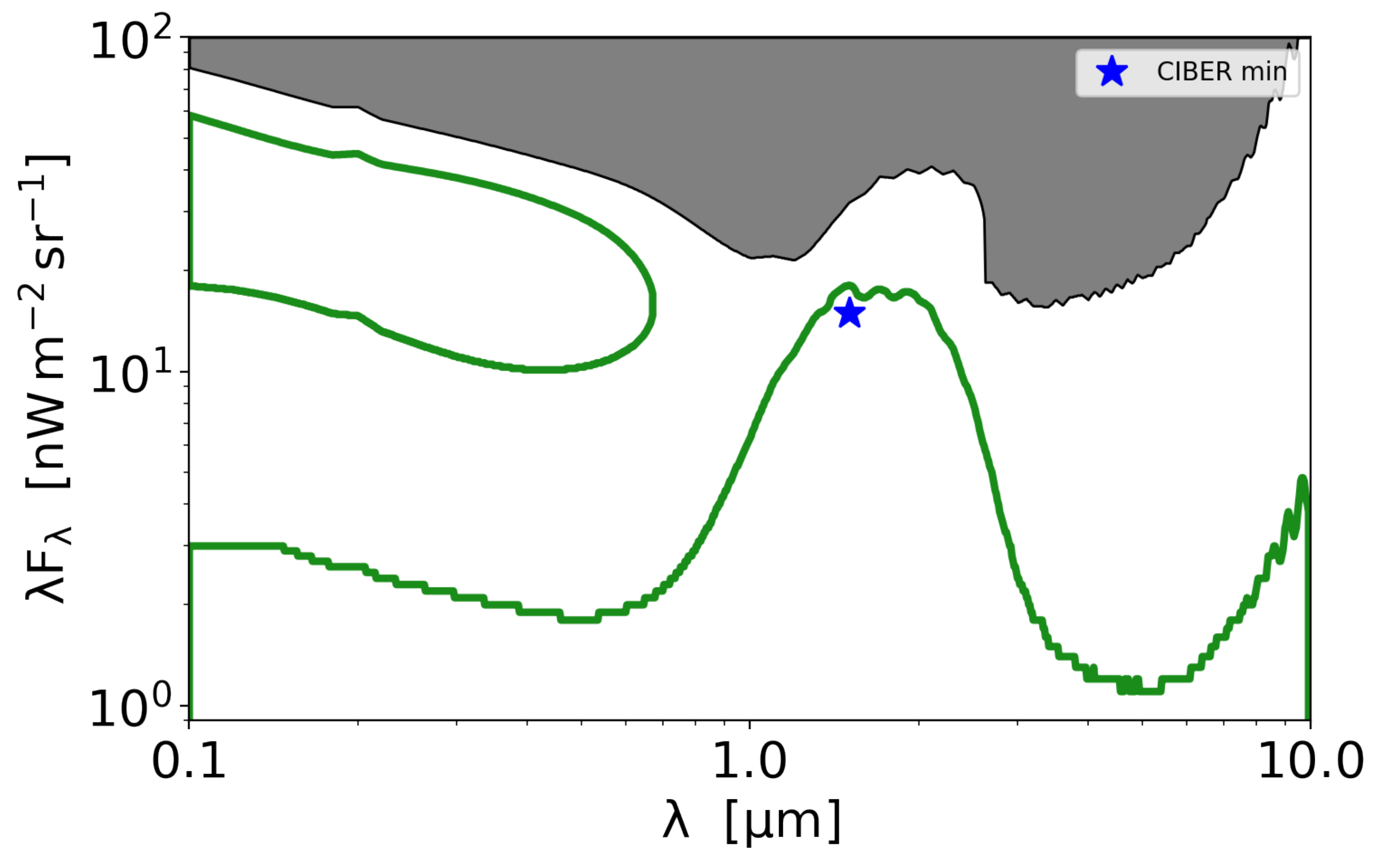}}
                \caption{Cross section of three dimensional parameter space calculated for the bump width $\sigma=0.2\ \mu$m. The cross sections of 95\% confidence level upper bound and the 68\% confidence level contour around the best-fit value for the normalisation of the log-Gaussian EBL bump as a function of the central wavelength are shown with grey and green correspondingly. The data of 1ES 1218+304 are used. Blue star shows the measurement of the minimal EBL bump by CIBER \citep{Matsuura:2017lub}.}
                \label{fig:ebl_bplaw1} 
        \end{figure}

The excess in the EBL spectrum is characterised by the following three parameters: central wavelength, width, and normalisation. In Fig. \ref{fig:ebl_bplaw} the central wavelength is fixed to show the dependence of constraint on normalisation as a function of the feature width. The source that imposes the strongest constraint is 1ES 1218+304. For this source, we show the dependence of the constraint on normalisation as a function of the central
wavelength in Fig. \ref{fig:ebl_bplaw1} . We fixed the width of the feature to $\sigma=0.2$~$\mu$m, which is equal to the width of the spectral feature modelled based on the CIBER minimal EBL measurements. 

The addition of the feature  at the central wavelength $\mu\simeq 1.7$~$\mu$m improves the quality of the fit if the normalisation is $B\simeq 15$~nW/(m$^2$ sr), compared to the nominal low EBL model. The best-fit 'bump' feature is shown superimposed on the low EBL model in Fig. \ref{fig:ebl}.  One could notice an agreement of the best fit bump feature with the measurements of CIBER. This agreement is also evident in Fig. \ref{fig:ebl_bplaw1} since the CIBER measurement is within the 68\% confidence contour of the \gr\ measurement.

We verified that the hypothesis of existence of the bump at $\simeq 1.7$ micron in the EBL spectrum does not also contradict the data on other blazars. This is demonstrated in Fig. \ref{fig:specs}, where fits of the spectra of all blazars considered in the analysis are shown for the EBL model with and without the 1.5 micron bumps with green and black thin lines, respectively. Table \ref{table:bl_params} provides details on the quality of the fits with different models.

\section{Discussion}

We have shown that the addition of a narrow spectral bump feature in the EBL at the level of minimal excess EBL reported by CIBER does not contradict the existing \gr\ data. On the other hand, the possibility of existence of stronger and wider excesses in the 0.1-10 $\mu$m EBL spectrum is strongly constrained by the \gr\ data. 

The narrow EBL bump produces a characteristic 'dip' feature in the \gr\ spectra shown in Fig. \ref{fig:specs}, which generally improves the quality of the fit. Hints of such dips are visible in the spectra of PKS 2155-304, PKS 2005-489, and in 1ES 1218+304, but not at the level of significance that is sufficient for detection. The dip features could also arise if the EBL has sharply falling intensity between the near-IR to mid-IR \citep{Orr:2011}. This model will lead to a slightly different shape in the attenuation of the gamma-ray flux as a function of energy and it should be possible to distinguish between the two possibilities once the statistics of the spectral measurements and control over systematics improve.

The effect of absorption of TeV \gr s in interactions with the EBL bump photons increases proportionally to the distance of the source. Therefore the TeV range dip impinged on the \gr spectrum by the EBL bump should get more pronounced for further away sources. However, the overall suppression of the spectrum in the multi-TeV range makes the quality of the spectra that are collected with the current generation of telescopes insufficient for the measurement of the effect of the EBL bump on the spectra. 

The situation will be improved upon the start of operations of the Cherenkov Telescope Array (CTA). For nearby blazars, such as PKS 2155-304, PKS 2005-489 and 1ES 1218+304, CTA will be able to provide high signal-to-noise measurements of the spectrum over sufficiently large dynamic ranges from tens of GeV up to 10 TeV. This will allow one to confirm or reject the hint of existence of the TeV dip in the \gr\ spectra (corresponding to the $\mu$m bump in the EBL spectrum with high significance). For further away sources, CTA will extend the dynamical range of the spectral measurements in the multi-TeV range.\ This would enable the confirmation and rejection of the EBL bump origin of the TeV dip in the spectra, that is, if the dip were to be systematically detected in further away sources with larger amplitude. 

If the dip in the 1ES 1218+304 spectrum is really due to the presence of a bump in the EBL spectrum, the improvement of the quality of spectral measurements with CTA will lead to the discovery of the EBL spectral feature.  We made an estimate of significance of detection that could be attained  by CTA. To do  this, we took current measurements of 1ES 1218+304 spectrum by VERITAS and divided the errorbars by the square root of the ratio of effective areas of CTA and VERITAS, which is by a factor of 3 (energy dependent) in the 0.3-3 TeV energy range. In this way we obtained an expected CTA-level quality of spectral measurements for the same exposure as in VERITAS ($\sim$ 86 hours). We excluded the first two points of the spectrum that lie below the model. With such a 'pre-fabricated' CTA-quality spectrum, we repeated the analysis described in the previous sections and found that  CTA will be able to help discover the spectral feature in the EBL spectrum and detect it with significance above $5\sigma$. Bump-like features in the EBL spectrum could be produced by decaying particles \citep{kohri17}, such as axion-like particles that possibly form part of dark matter \citep{axions}, and by peculiar stellar populations that could have existed in the past but are absent today, such as Population III stars \citep{populationIII}. In this respect, the possibility of discovering spectral features in the EBL spectrum via precision measurements of the blazar spectra that would be enabled by CTA will be relevant in a range of astrophysics and cosmology in addition to fundamental physics contexts.

\section*{Acknowledgements}

The work of AK on the analysis of gamma-ray spectra 
and EBL modelling was supported by the
Russian Science Foundation, grant 18-12-00258. 
AK's stay in the APC laboratory was provided by the 
scholarship 'Vernadsky' of the French embassy in Russia.

\bibliographystyle{aa}
\bibliography{mybib}

\end{document}